\documentclass[acmsmall,screen,nonacm]{acmart}

\newif\ifarxiv
\arxivtrue

\usepackage{booktabs}
\usepackage{multirow}
\usepackage{xspace}
\usepackage{graphicx}
\usepackage{amsmath}
\usepackage{tabularx}
\usepackage{placeins}

\newcommand{\system}{\textsc{AgentSecGraph}\xspace}
\newcommand{\benchmark}{\textsc{AgentSecBench}\xspace}
\newcommand{\securityadg}{\textsc{Security-ADG}\xspace}

\renewcommand\footnotetextcopyrightpermission[1]{}
\begin{document}

\title[Beyond Predefined Sinks]{Beyond Predefined Sinks:
Security-Aware Dependency Analysis for LLM Agents}

% TODO(arXiv): Confirm the complete author order, affiliations, and ORCIDs.
\author{Hang Cui}
\email{cuihang26@mails.ucas.ac.cn}
\renewcommand{\shortauthors}{Hang Cui}

\begin{abstract}
Large language model (LLM)-based agents increasingly connect model-generated
decisions to security-sensitive software capabilities, including command
execution, filesystem access, network communication, browser control, and
external tools. Existing analyses often use predefined sensitive operations
to anchor their analysis, but operation identity alone is insufficient to
determine the security implications of an agent behavior.

We present \system, a security-aware static analysis framework that constructs
a candidate-centered Security-Aware Agent Dependency Graph (\securityadg) for
each security-sensitive operation. The representation augments operation
identity with agent relevance, source and dependency evidence, trust-boundary
context, guard evidence, and external-effect semantics.

We further introduce \benchmark, a reproducible corpus of 67 real-world
LLM-agent repositories spanning 11 agent ecosystems and 37,542 source files.
Applying the current analyzer to all frozen repositories yields 23,866 static
security-sensitive operation candidates across 65 repositories, for which
\system emits 23,866 candidate-centered \securityadg artifacts without
reported construction failures. Corpus-wide analysis recovers
source-to-operation dependency evidence for 9,821 candidates (41.15\%) and
potential guard evidence for 3,075 (12.88\%). The complete analysis finishes
in 50.8 minutes.

Using a separate reproduction-backed evaluation layer, we establish a
conservative reference set of 22 security-sensitive behaviors across 13
repositories: one confirmed vulnerability, one pending disclosure candidate,
and 20 guarded behaviors that do not meet the vulnerability criteria. In a
held-out representation study, \securityadg preserves 91.1\% of the reference
context and all five observed guards, compared with 20.0\% for a sink-only
view and 40.0\% for a simplified ADG.

These results show that security-aware dependency and contextual evidence
enable distinctions that cannot be recovered from sensitive-operation
identity alone.
\end{abstract}


\maketitle

\hypersetup{
  pdftitle={Beyond Predefined Sinks: Security-Aware Dependency Analysis for LLM Agents},
  pdfauthor={Hang Cui},
  pdfsubject={Security analysis of LLM agents},
  pdfkeywords={LLM agents, static analysis, security, dependency graphs}
}
\pdfinfo{
  /Author (Hang Cui)
  /Keywords (LLM agents, static analysis, security, dependency graphs)
}

\begin{center}
\small Preprint. This version has not been peer reviewed.
\end{center}

\section{Introduction}
\label{sec:introduction}

Large language model (LLM)-based agents are rapidly evolving from
text-generating assistants into software systems capable of planning,
invoking tools, maintaining state, and interacting with external
environments. Early paradigms such as ReAct~\cite{react2023} and
Toolformer~\cite{toolformer2023} demonstrated how language models can
interleave reasoning with external actions. Modern frameworks, including
LangChain~\cite{langchain2023}, LangGraph~\cite{langgraph2024},
AutoGen~\cite{autogen2024}, and CrewAI~\cite{crewai2024}, provide
reusable abstractions for tool invocation, memory, workflow orchestration,
and multi-agent coordination. The Model Context Protocol
(MCP)~\cite{mcp2024} extends this ecosystem by standardizing access to
external tools and resources.

This transition fundamentally changes the security boundary of LLM
applications. In a conventional chatbot, an unsafe model output may remain
textual; in an agent system, the same output can influence command
execution, filesystem operations, browser actions, database access, or
remote services. Agent behavior may additionally depend on user
instructions, retrieved documents, memory contents, model-generated plans,
tool outputs, and messages from other agents. Untrusted information can
therefore cross multiple software and trust boundaries before affecting an
externally visible operation. Prior work has demonstrated these risks through indirect
prompt injection~\cite{greshake2023indirect,injecagent2024}, realistic
agent attack environments~\cite{agentdojo2024}, unsafe tool
interactions~\cite{toolemu2024}, agent security
benchmarks~\cite{asb2025}, resource abuse~\cite{agentdos2026}, and
MCP-specific attack surfaces~\cite{mcplandscape2026,mcpsafetybench2026}.

Static program analysis provides a natural foundation for reasoning about these
behaviors. Source--sink and taint analyses determine whether values from
selected sources may reach security-sensitive operations. Systems such as
YASA~\cite{yasa2026} demonstrate scalable taint analysis across
multiple languages, while recent work has adapted program analysis to
LLM-integrated applications. AgentFuzz~\cite{agentfuzz2025} dynamically
searches for prompts that trigger taint-style vulnerabilities,
TaintP2X~\cite{taintp2x2026} statically traces LLM-generated values toward
sensitive operations, and Agent Audit~\cite{agentaudit2026} combines
agent-aware data-flow reasoning with credential, configuration, and
privilege analysis.

However, \emph{operation identity alone is insufficient for security
interpretation}. The same security-sensitive operation may be directly
influenced by an agent-facing input, disconnected from that input by program
transformations, or constrained by validation and authorization logic.
Conversely, security-relevant capabilities may be exposed through
framework-specific abstractions rather than obvious low-level API calls.
A useful analysis must therefore recover not only \emph{which} operation is
present, but also \emph{how} agent-relevant information reaches it and the
\emph{security context in which} it executes.

\begin{figure*}[t]
    \centering
    \includegraphics[width=0.94\textwidth]{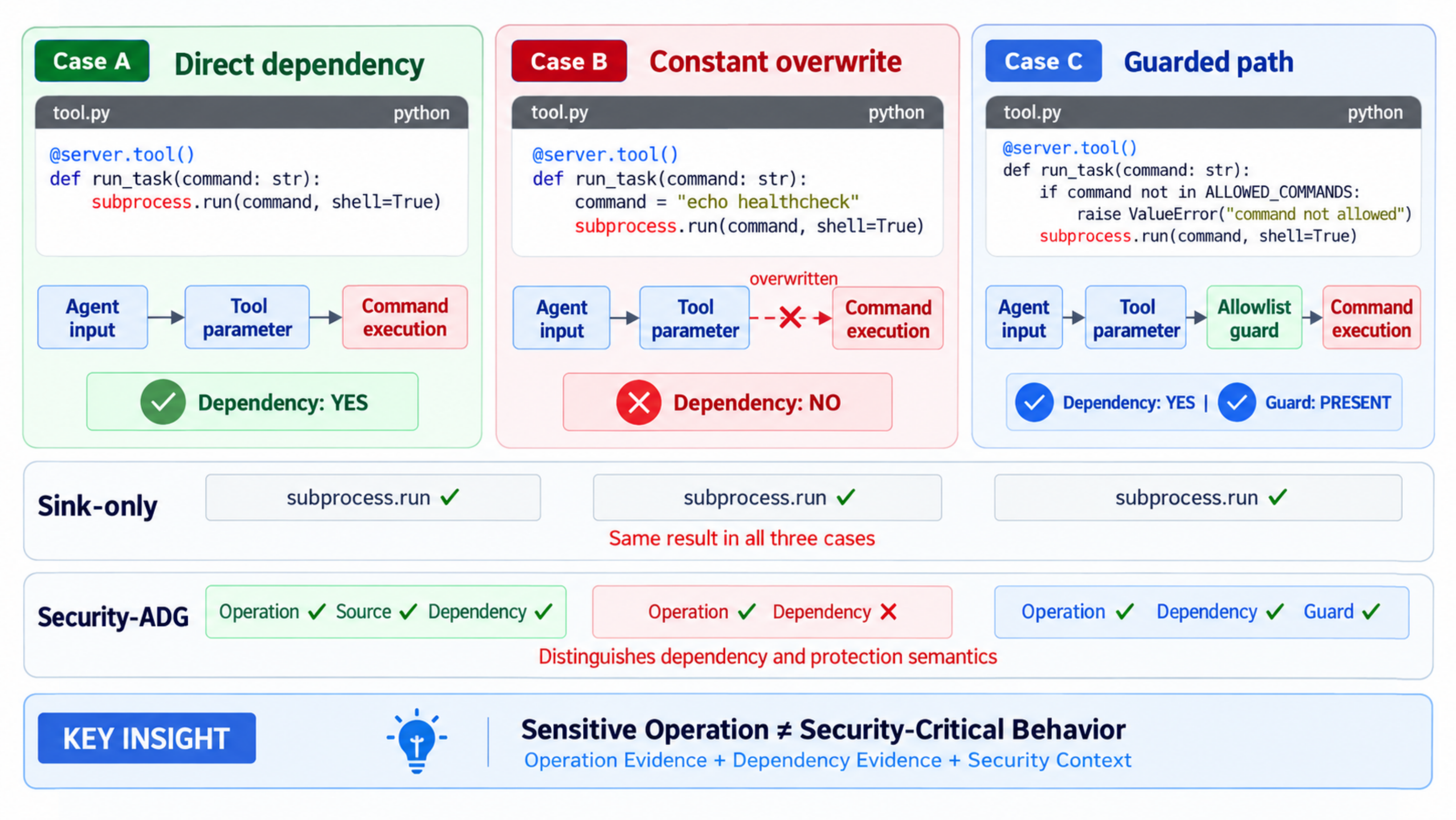}
    \caption{\textbf{Motivation: operation identity alone does not determine
    security semantics.}
    All three agent tools contain the same command-execution operation.
    Cases A and C preserve a dependency from the agent-facing input,
    whereas Case B removes that dependency through a constant overwrite.
    Case C additionally contains a candidate allowlist guard.
    Security-aware dependency analysis distinguishes these cases using
    dependency and security-context evidence.}
    \label{fig:motivation}
\end{figure*}

Figure~\ref{fig:motivation} illustrates this distinction.

Agent frameworks further complicate this problem because relevant
relationships may be hidden behind decorators, tool wrappers, workflow
nodes, dynamically registered functions, plugins, and remote MCP tools.
AgentFlow~\cite{agentflow2026}, for example, recovers framework-induced
Agent Dependency Graphs (ADGs) over prompts, models, capabilities, memory,
and control policies. Recovering these structural dependencies is valuable,
but security interpretation also requires evidence about source
provenance, trust boundaries, guards, privilege context, and resulting
external effects.

This motivates the central research question of this work:

\begin{quote}
\textit{How can static analysis augment security-sensitive operations with
agent-specific dependency and security-context evidence to analyze
security-critical behaviors in real-world LLM-agent programs?}
\end{quote}

We present \textbf{\system}, a security-aware static analysis framework for
LLM-agent programs. Rather than treating a matched sensitive operation as a
vulnerability verdict, \system uses it as an analysis anchor and constructs
a candidate-centered \textbf{\securityadg}. The representation augments
operation identity with agent relevance, source provenance, local dependency
evidence, trust-boundary context, candidate guards, statically available
privilege information, and external-effect semantics. This design
separates three analytical layers:

\begin{equation}
\text{Operation}
\rightarrow
\text{Dependency}
\rightarrow
\text{Security Context}.
\label{eq:intro-evidence-layers}
\end{equation}

This separation allows \system to distinguish behaviors that appear
identical under an operation-only view while making the limitations of the
underlying static evidence explicit.

We further introduce \textbf{\benchmark}, a reproducible corpus of
67 real-world LLM-agent repositories spanning 11 agent ecosystems,
37,542 source files, and 291.2\,MB of source code. Applying the current
candidate extractor to the full corpus identifies 23,866 static
security-sensitive operation candidates across 65 repositories and 7,214
candidate-bearing source files. \system emits one candidate-centered
\securityadg artifact for every candidate without reported construction
failures. Across this population, \system recovers
source-to-operation dependency evidence for 9,821 candidates (41.15\%),
explicit framework-specific evidence for 749 (3.14\%), and potential guard
evidence for 3,075 (12.88\%). End-to-end analysis completes in 50.8 minutes.

Separately, we retain a frozen pilot cohort and construct
reproduction-backed evidence for 22 security-sensitive behaviors across
13 repositories. The resulting reference set contains one confirmed
vulnerability, one pending disclosure candidate, and 20 guarded
non-vulnerabilities under the evaluated conditions; neither model-generated
judgments nor single-reviewer labels are used as ground truth. Across nine
evidence-backed held-out cases, \securityadg preserves 91.1\% of the
reference context and all five observed guards, compared with 20.0\%
context completeness and 0/5 guard visibility for a sink-only view, and
40.0\% and 0/5, respectively, for a simplified ADG view.

Our contributions are summarized as follows:

\begin{itemize}

    \item \textbf{Security-analysis formulation.}
    We formulate LLM-agent security analysis by distinguishing among
    sensitive-operation identity, security-critical behavior evidence, and
    vulnerability claims, avoiding the assumption that every matched
    sensitive operation constitutes a vulnerability.

    \item \textbf{Security-aware dependency analysis.}
    We design \system and its candidate-centered \securityadg
    representation to integrate agent relevance, source provenance, local
    dependency evidence, trust-boundary context, candidate guards, and
    external-effect semantics.

    \item \textbf{Benchmark and corpus-wide analysis.}
    We construct \benchmark from 67 frozen real-world LLM-agent repositories
    spanning 11 ecosystems and 37,542 source files. A full-corpus run of the
    current analyzer emits 23,866 candidate-centered \securityadg artifacts
    across 65 repositories, enabling corpus-wide characterization
    and scalability analysis.

    \item \textbf{Reproduction-backed empirical evidence.}
    We establish a reproduction-confirmed reference set of 22
    security-sensitive behaviors across 13 repositories and quantify the
    security-context information preserved by sink-only, simplified ADG,
    and full \securityadg representations.

\end{itemize}
\section{Motivation and Problem Formulation}
\label{sec:background}
\label{sec:problem}

LLM-agent software differs from conventional language-model applications:
model-mediated decisions can directly trigger executable capabilities.
Modern agents combine tools, memory, retrieval, workflows, and multi-agent communication
~\cite{react2023,toolformer2023,langchain2023,langgraph2024,
autogen2024,crewai2024}, while protocols such as MCP~\cite{mcp2024}
standardize access to external tools and resources.

These capabilities introduce security-relevant relationships that are often
hidden behind framework abstractions such as tool decorators, workflow
nodes, dynamically registered functions, plugins, and inter-agent handoffs.
Consequently, security analysis must reason not only about host-language
operations, but also about how agent-facing inputs can access and influence
those operations.

\subsection{Agent-Specific Security Semantics}
\label{sec:agent-semantics}

Untrusted information in an agent application may originate from user
instructions, retrieved documents, web content, memory, model outputs, or
external tools. Such values may subsequently influence command execution,
filesystem access, network communication, browser actions, database
operations, or other externally visible effects.

Static source--sink and taint analysis provides a natural foundation for
recovering these relationships. Classical program representations such as
Program Dependence Graphs (PDGs)~\cite{pdg1987} capture data and control
dependencies, while systems such as FlowDroid~\cite{flowdroid2014} and
YASA~\cite{yasa2026} demonstrate precise and scalable source--sink
reasoning. Recent work extends these ideas to LLM-integrated software:
TaintP2X~\cite{taintp2x2026} traces LLM-generated values toward sensitive
operations, Agent Audit~\cite{agentaudit2026} incorporates agent-specific
configuration and privilege information, and AgentFuzz~\cite{agentfuzz2025}
dynamically searches for inputs that trigger sensitive behavior.

However, LLM-agent software introduces an additional semantic layer.
A parameter exposed through an agent tool interface has different security
implications from an unrelated internal parameter, even if both eventually
reach the same low-level API. Framework-induced relationships must therefore
be interpreted alongside ordinary program dependencies.

\subsection{Why Sink Identity Is Insufficient}
\label{sec:sink-insufficient}

Figure~\ref{fig:motivation} illustrates the central limitation of
operation-only analysis. All three examples contain the same
command-execution operation, but their security semantics differ:
one preserves a direct dependency from an agent-facing input, one removes
that dependency through a constant overwrite, and one retains the dependency
while introducing an allowlist guard.

This observation motivates a distinction among three levels of evidence.
A \emph{security-sensitive operation candidate} is a source-code operation
that may produce or prepare a security-relevant external effect.
A \emph{security-critical behavior} combines such an operation with
agent-specific dependency and contextual evidence. A
\emph{vulnerability} is a stronger conclusion that additionally depends on
factors such as controllability, protection mechanisms, deployment exposure,
authorization, privilege, impact, and exploitability.

Accordingly,

\begin{equation}
\text{Sensitive Operation}
\neq
\text{Security-Critical Behavior}
\neq
\text{Vulnerability}.
\label{eq:behavior-distinction}
\end{equation}

Predefined operation specifications are therefore useful for locating code
that deserves further analysis, but operation identity should be treated as
an \emph{analysis anchor}, not as a vulnerability verdict.

\subsection{Security Context and Analysis Objective}
\label{sec:security-context-objective}

Let

\begin{equation}
\mathcal{O}=\{o_1,o_2,\ldots,o_n\}
\label{eq:candidate-set}
\end{equation}

denote the set of security-sensitive operation candidates in a frozen
program snapshot. Candidate generation is intentionally permissive: a match
does not imply that the operation is agent-controlled, reachable from an
untrusted source, insufficiently guarded, or exploitable.

For each candidate $o\in\mathcal{O}$, we characterize the statically
available security context as

\begin{equation}
C(o)=\langle R,S,D,T,P,G,E\rangle,
\label{eq:security-context}
\end{equation}

where $R$ denotes agent relevance; $S$, source provenance; $D$,
source-to-operation dependency evidence; $T$, trust-boundary evidence;
$P$, statically available privilege context; $G$, candidate guard evidence;
and $E$, the external-effect category.

These dimensions represent evidence rather than security verdicts.
Dependency or trust-boundary evidence does not establish exploitability,
while the presence of a guard does not establish that the guard is complete
or effective.

Given a frozen LLM-agent program and candidate set $\mathcal{O}$, our goal
is therefore to recover the dependency and security-context evidence needed
to interpret each candidate:

\begin{equation}
\text{Operation}
+
\text{Dependency}
+
\text{Security Context}
\rightarrow
\text{Behavior Evidence}.
\label{eq:analysis-objective}
\end{equation}

The objective is deliberately narrower than automatic vulnerability
verification. \system recovers auditable evidence about how an
agent-relevant value may influence a sensitive capability and what security
context surrounds that relationship. Any stronger vulnerability
determination is made separately using reproduction and security evidence.

Section~\ref{sec:approach} presents the resulting candidate-centered
\securityadg analysis.
\section{AgentSecGraph}
\label{sec:approach}

\subsection{Overview}
\label{sec:approach-overview}

\system is a static analysis framework for recovering auditable
security-relevant evidence associated with sensitive operations in LLM-agent
software. Rather than interpreting a matched operation as a security
violation, \system treats it as an analysis anchor and builds a
candidate-centered \securityadg for it.

Given a candidate operation $o\in\mathcal{O}$, \system determines how the
operation is exposed to agents, recovers relevant source-to-operation
dependencies,
extracts trust-boundary and guard evidence, and summarizes the resulting
security context using $C(o)$ from Equation~\ref{eq:security-context}.

\begin{figure*}[t]
    \centering
    \includegraphics[width=0.92\textwidth]{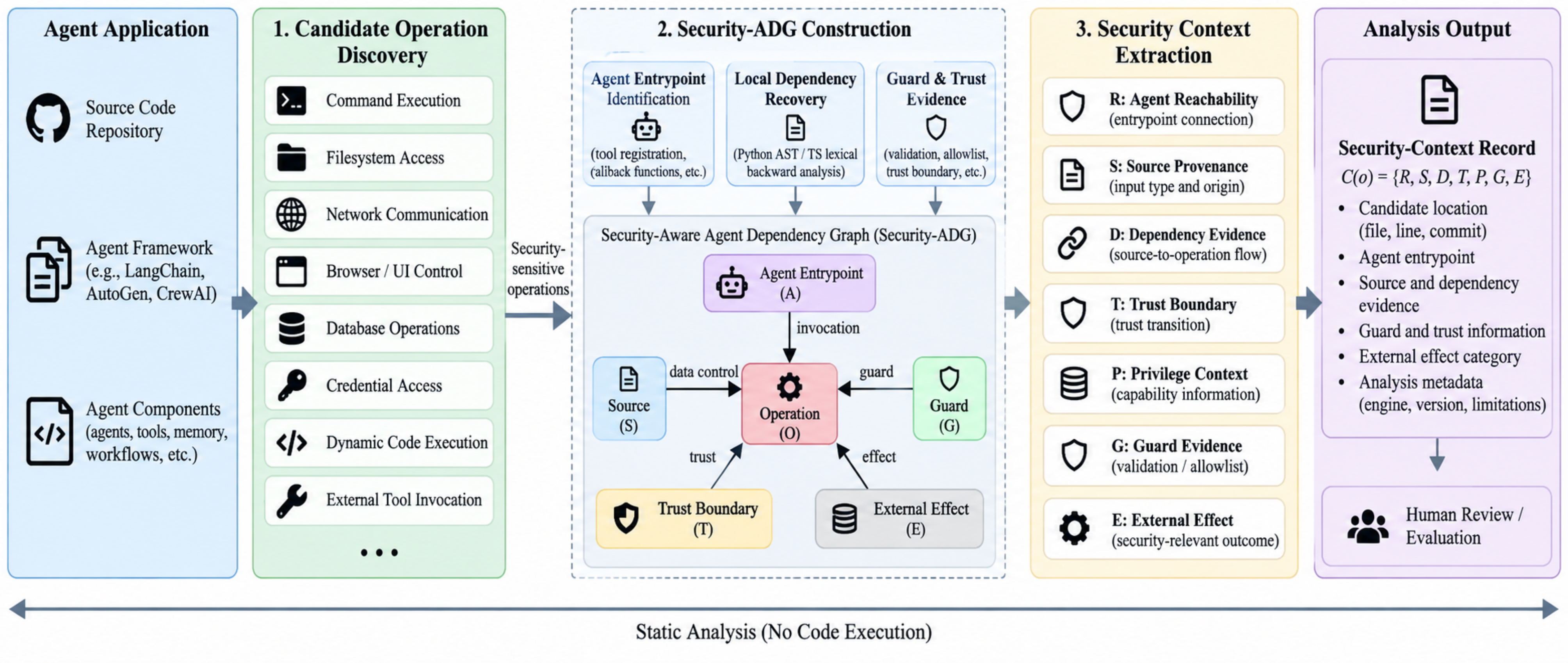}
    \caption{\textbf{Overview of \system.}
    Starting from a security-sensitive operation candidate, \system
    normalizes framework-specific agent semantics, identifies relevant
    sources, recovers dependency and security-context evidence, and
    constructs a candidate-centered \securityadg for subsequent analysis
    and review.}
    \Description{Overview of the AgentSecGraph analysis pipeline.
    A security-sensitive operation serves as the analysis anchor.
    Framework-specific agent semantics and input sources are normalized,
    local dependency, trust-boundary, guard, and external-effect evidence
    are recovered, and the resulting evidence is represented as a
    candidate-centered Security-ADG.}
    \label{fig:overview}
\end{figure*}

Figure~\ref{fig:overview} summarizes the end-to-end workflow.

The framework deliberately separates static evidence recovery from
vulnerability assessment. Its output records only what can be established
from the frozen source snapshot and retains uncertainty when the conclusion
depends on runtime, deployment, or threat-model evidence.

\subsection{Candidate Anchoring and Framework Normalization}
\label{sec:candidate-anchoring}
\label{sec:framework-normalization}

The analysis begins with security-sensitive operation candidates
$o\in\mathcal{O}$. Each candidate is associated with its frozen repository
snapshot, source location, operation category, detector provenance, and
local code context.

Candidate generation is intentionally permissive. A match establishes that a
potentially security-relevant capability exists, but not that the capability
is agent-controlled, reachable from an untrusted source, insufficiently
guarded, or exploitable.

Agent frameworks further complicate source-level interpretation because tools and
capabilities are exposed through heterogeneous abstractions, including
decorators, registration APIs, factories, workflow nodes, and tool classes.
\system therefore maps supported framework constructs into a common semantic
record

\begin{equation}
\mathcal{N}
=
\langle F,R_s,Y,P_s,Q\rangle,
\label{eq:framework-normalization}
\end{equation}

where $F$ identifies the framework; $R_s$, the semantic role of the
construct; $Y$, the associated program symbol; $P_s$, its exposed parameters;
and $Q$, the match confidence.

Normalization affects semantic interpretation rather than vulnerability
assessment. For example, when a function is recognized as an agent-facing
tool, its exposed parameters can be classified as agent-relevant sources
rather than ordinary function parameters.

The current implementation supports representative constructs from
MCP/FastMCP, LangChain, CrewAI, AutoGen, OpenAI Agents SDK,
Semantic Kernel, and LlamaIndex. The supplementary material documents the
normalization contract, supported construct families, and matching rules.

\subsection{Security-ADG Construction}
\label{sec:security-adg-construction}

For each candidate operation $o$, \system constructs a typed
Security-Aware Agent Dependency Graph

\begin{equation}
G_o=(V_o,E_o,\Phi_o),
\label{eq:security-adg}
\end{equation}

where $V_o$ contains evidence nodes associated with $o$, $E_o$ contains
recovered evidence relationships, and $\Phi_o$ stores graph-level provenance
and security-context attributes.

Unlike a whole-program dependency graph, $G_o$ is candidate-centered.
Its purpose is not to reconstruct every program relationship, but to retain
the evidence required to interpret one security-sensitive operation.

\begin{figure*}[t]
    \centering
    \includegraphics[width=0.90\textwidth]
        {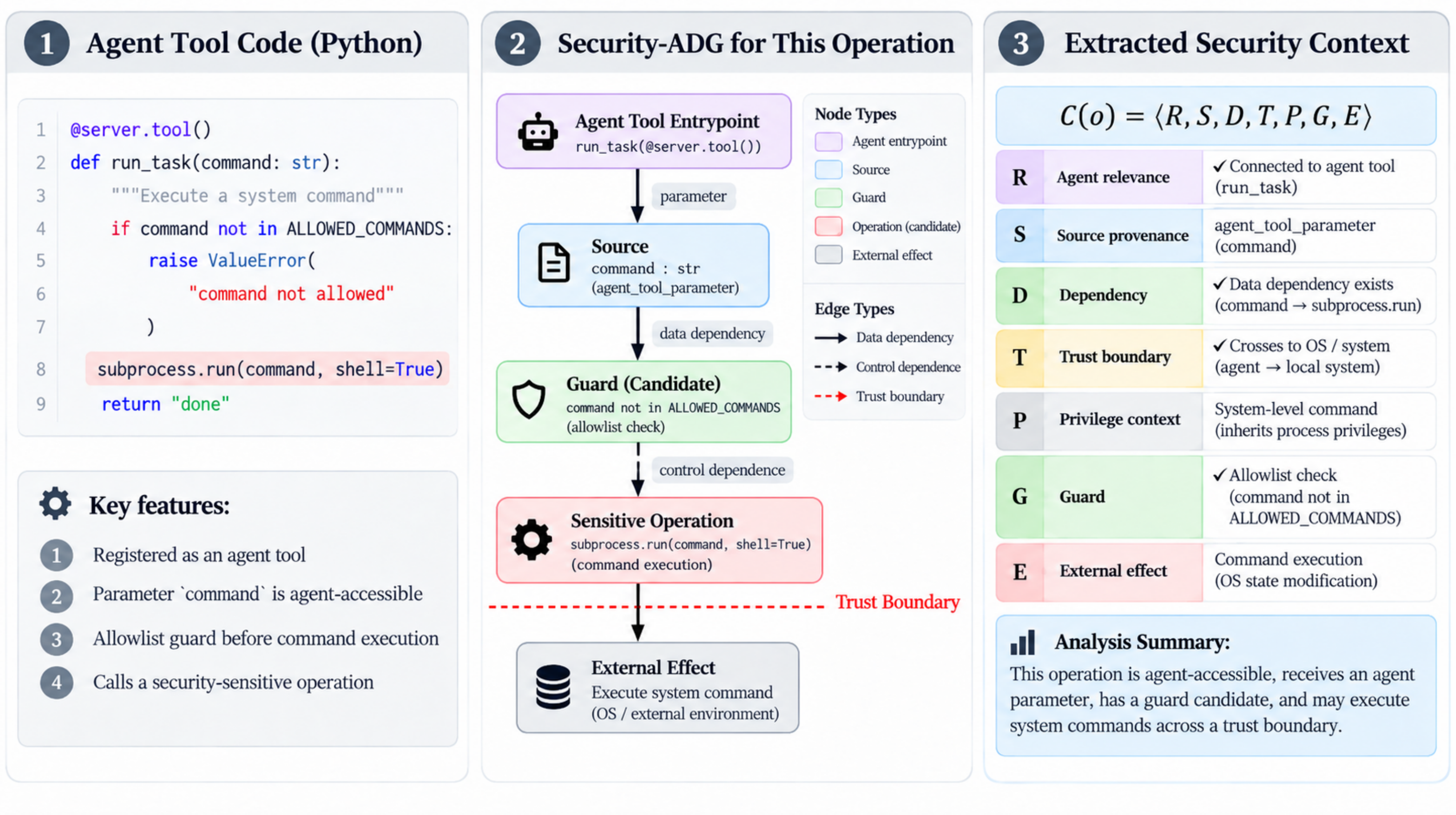}
    \caption{\textbf{Example candidate-centered \securityadg.}
    The representation augments a security-sensitive operation with
    agent-facing source provenance, dependency relationships,
    trust-boundary evidence, candidate guards, and the resulting external
    effect. Relations prefixed with \texttt{may\_} represent static
    evidence rather than proofs of reachability or exploitability.}
    \Description{Example Security-Aware Agent Dependency Graph centered on
    a security-sensitive operation. The graph contains an agent-facing
    program construct, input source, dependency evidence, trust boundary,
    guard evidence, sensitive operation, and external effect.}
    \label{fig:security-adg-example}
\end{figure*}

Figure~\ref{fig:security-adg-example} shows a representative
candidate-centered \securityadg.

Table~\ref{tab:security-adg-schema} summarizes the core node and edge types.

\begin{table}[t]
    \caption{Core node and edge types in \securityadg. Relations prefixed
    with \texttt{may\_} represent static evidence rather than reachability
    or exploitability proofs.}
    \label{tab:security-adg-schema}
    \centering
    \small
    \setlength{\tabcolsep}{5pt}
    \renewcommand{\arraystretch}{1.08}
    \begin{tabularx}{\linewidth}{@{}lX@{}}
        \toprule
        \textbf{Type} & \textbf{Meaning} \\
        \midrule

        Agent/program symbol &
        Agent-facing or enclosing program construct associated with the
        candidate. \\

        Sensitive operation &
        Candidate operation that performs or prepares a security-relevant
        effect. \\

        Input source &
        Value originating from an agent-facing or external context. \\

        Trust boundary &
        A crossing from a less-trusted or external context to a local
        capability. \\

        Guard candidate &
        Validation, authorization, sandboxing, argument restriction, or
        related defensive evidence. \\

        External effect &
        An abstract effect involving a process, filesystem, network, browser,
        or external service, deterministically assigned from the operation
        category. \\

        \midrule

        \texttt{contains} &
        A program symbol contains the candidate operation. \\

        \texttt{may\_cause} &
        The candidate may produce the represented external effect. \\

        \texttt{may\_data\_depend\_on} &
        A source may influence values consumed by the candidate. \\

        \texttt{may\_cross} &
        Evidence suggests that a trust boundary is crossed. \\

        \texttt{may\_guard} &
        A guard candidate may constrain the operation. \\

        \bottomrule
    \end{tabularx}
\end{table}

The use of \texttt{may\_*} relations is deliberate. These edges represent
static evidence and possible relationships; they are not proofs of
end-to-end reachability, attacker control, exploitability, or guard
sufficiency.

\subsection{Dependency, Trust-Boundary, and Guard Analysis}
\label{sec:dependency-recovery}
\label{sec:guard-trust-analysis}
\label{sec:context-extraction}

Starting from values consumed by a candidate operation, \system performs
backward dependency recovery to identify whether a recognized source may
influence the candidate. Conceptually, the analysis seeks a local chain

\begin{equation}
s
\rightsquigarrow
v_1
\rightsquigarrow
\cdots
\rightsquigarrow
v_o
\rightarrow
o,
\label{eq:dependency-path}
\end{equation}

where $s$ is an agent-relevant or externally influenced source and $v_o$ is
consumed by operation $o$.

For Python, dependency recovery is AST-based and follows local parameters,
assignments, definitions, and reassignments. For example, if an agent-facing
parameter is overwritten by a constant before reaching the candidate, the
corresponding dependency is terminated.

For TypeScript and JavaScript, the current implementation performs
syntax-aware lexical backward tracing over local parameters, definitions,
assignments, and supported registration patterns. This analysis is more
conservative than the Python engine and does not provide complete alias,
type, or control-flow reasoning.

Dependency evidence alone does not imply that a behavior is unsafe.
\system therefore also records evidence of safeguards and trust boundaries.
Representative guard evidence includes allowlists, fixed-argument or
argument-vector command construction, schema or syntax validation,
authentication checks,
path restrictions, sandboxing, and resource limits.

A \texttt{may\_guard} edge indicates only that defensive evidence is visible;
it does not prove that the guard is complete or effective. Similarly,
\texttt{may\_cross} records available evidence that an external,
agent-facing, or otherwise less-trusted value may influence a local or more
privileged capability, while deployment-dependent trust and privilege may
remain unknown.

The current implementation represents privilege context only when explicit
evidence is recoverable from supported source-level constructs. It does not
infer deployment identities, ambient permissions, or runtime authorization
state. Likewise, every candidate is deterministically mapped to an abstract
external-effect category. The resulting effect node encodes modeled
semantics rather than a runtime-confirmed effect and is therefore a schema
property, not an empirical recovery result.

The recovered evidence is summarized using the security-context tuple
$C(o)$ defined in Equation~\ref{eq:security-context}, whereas the graph itself
preserves source-level provenance and relationships for inspection.
In particular, $D(o)=\mathrm{true}$ means that dependency evidence was
recovered; it does not imply
$\mathrm{Vulnerability}(o)=\mathrm{true}$.

\subsection{Implementation and Analysis Contract}
\label{sec:implementation}
\label{sec:implementation-engines}
\label{sec:implementation-adapters}
\label{sec:implementation-artifacts}
\label{sec:representation-projections}
\label{sec:analysis-contract}

\system is implemented as a modular source-analysis pipeline with
language-specific analyzers, framework adapters, and deterministic graph
generation.

The Python engine uses the language AST to recover enclosing scopes,
parameters, local definitions, assignments, decorators, and backward
def--use relationships. The TypeScript/JavaScript engine performs
syntax-aware lexical tracing and explicitly records the weaker guarantees
provided by this approach.

Framework adapters map framework-specific tool declarations and
capability-registration mechanisms into the normalized representation from
Equation~\ref{eq:framework-normalization}. The adapter layer is independent
of the graph schema, allowing new frameworks to be supported without
redesigning \securityadg.

For each candidate, the implementation emits a candidate-centered
\securityadg together with source-level provenance and analysis metadata.
We report graph construction as successful when the deterministic pipeline
emits exactly one graph artifact linked to the original candidate identifier
and source provenance. This measures one-to-one generation and linkage, not
the semantic correctness or completeness of recovered relationships.
It also derives three deterministic representation views from the same
candidate:

\begin{enumerate}
    \item \emph{sink-only}, retaining only the sensitive operation;
    \item \emph{simplified ADG}, retaining structural operation and
          external-effect information; and
    \item full \securityadg, retaining source/dependency,
          trust-boundary, guard, and external-effect evidence.
\end{enumerate}

These projections allow representation-level comparisons without changing
the underlying candidate population.

All analyses operate on repositories frozen at recorded Git commits.
Candidate-generation rules, framework adapters, dependency logic, and graph
construction are independent of human annotations, model-generated security
judgments, reproduction outcomes, and vulnerability dispositions.

The current analysis is intentionally local. Python analysis is not a
complete whole-program interprocedural, alias, points-to, or symbolic
analysis, and TypeScript/JavaScript tracing does not provide sound, complete
alias or control-flow analysis. Reflection, dynamic registration,
metaprogramming, and deployment-dependent exposure or privilege may
therefore produce incomplete evidence.

Accordingly, \securityadg should be interpreted as an auditable static
evidence representation for a candidate operation, not as a proof of
reachability, exploitability, guard sufficiency, or vulnerability.
\section{Evaluation}
\label{sec:agentsecbench}
\label{sec:evaluation}
\label{sec:results}

We evaluate \system on frozen snapshots of real-world LLM-agent repositories.
Our four questions concern corpus-wide coverage and scalability,
conservative behavior-level ground truth, distinctions beyond sink identity,
and the security context preserved by \securityadg.

Throughout the evaluation, we distinguish three levels of evidence:
\emph{static candidates}, \emph{reproduction-confirmed behaviors}, and
\emph{vulnerability dispositions}. Static candidates are analysis inputs
rather than vulnerability labels.

\subsection{Evaluation Populations and Experimental Setup}
\label{sec:corpus-construction}
\label{sec:corpus-characteristics}
\label{sec:corpus-operations}

\paragraph{Corpus construction.}
\benchmark is a corpus of publicly available LLM-agent repositories drawn
from diverse ecosystems, including MCP-based systems, coding agents,
framework-based applications, autonomous agents, and multi-agent platforms.
We retain repositories containing substantive agent logic and analyzable
source code while excluding duplicates, mirrors, trivial examples, and
repositories unsuitable for reproducible source-level analysis.

Repository inclusion is independent of \system output. Each retained
repository is frozen at a recorded Git commit so that candidate extraction,
graph construction, reproduction, and evaluation refer to the same source
snapshot. The supplementary material documents the collection criteria and
aggregate corpus composition. \ifarxiv The public artifact repository provides
\else The anonymous replication package will provide \fi
{}the complete row-level repository inventory and immutable commit identifiers.

Table~\ref{tab:evaluation-overview} summarizes the corpus-wide analysis and
the reproduction-confirmed reference set.

Python is the dominant implementation language: 51 of 67 repositories
(76.1\%) primarily use Python, while 7 (10.4\%) primarily use TypeScript.
The remaining repositories are implemented mainly in C\#, Rust, Go, PHP,
and Shell. This imbalance is relevant because \system provides stronger
AST-based dependency recovery for Python than the lexical analysis currently
used for TypeScript/JavaScript.

\paragraph{Evaluation populations.}
We distinguish three candidate populations throughout the evaluation. The
\emph{frozen pilot cohort} contains 531 candidates from five repositories
and is retained for compatibility with earlier pilot and annotation
studies. Running the current scanner on the same five repository snapshots
yields 508 candidates because later language scoping and AST-based filtering
changed candidate extraction. We call this the \emph{current
five-repository scan} and use it only to document scanner-version drift; we
do not merge it with the frozen cohort. The \emph{current full-corpus run}
applies the current scanner to all 67 frozen repositories and yields 23,866
static security-sensitive operation candidates. This population is used for
corpus-wide characterization and scalability measurements. None of these
candidate counts denotes vulnerabilities.

\paragraph{Full-corpus candidate extraction.}
The candidate taxonomy includes command/process execution, dynamic code
execution, filesystem access, network and remote access, browser and desktop
automation, database operations, credential or secret access, and
external-tool invocation. Applying the current scanner to all 67 frozen
repositories yields 23,866 candidates from 7,214 source files. Sixty-five
repositories contain at least one candidate. All 67 repositories complete
without skips or scanner failures. These candidates form the population for
corpus-wide \securityadg characterization; they may represent legitimate
capabilities, disconnected operations, guarded behaviors, or potentially
unsafe paths.

\paragraph{Experimental environment.}
All experiments were conducted on an HP ProBook 440 14-inch G10 equipped with
a 13th Gen Intel Core i5-1340P CPU (16 logical processors) and 32\,GB of RAM,
running Microsoft Windows 11 Home (build 26200). The implementation used
CPython 3.13.12 from the Anaconda distribution. Candidate extraction and
\securityadg construction each used a single analysis process with no worker
pool; repository Git reads were issued sequentially. The frozen repositories
and generated artifacts were stored on a 1\,TB WD Elements SE USB SSD.
Wall-clock measurements are from one completed full-corpus run. The archived
aggregate records phase and per-repository timings, but not repeated-run
measurements or a formal peak-RSS measurement; we therefore report the
observed wall time rather than a mean or stable cross-system performance
estimate.

\paragraph{Freeze and leakage control.}
Each evaluation population is analyzed under a fixed configuration. The
historical 531-candidate pilot manifest is preserved unchanged, whereas the
current full-corpus analysis uses the later scanner version frozen before
corpus-wide characterization. Within the current full-corpus evaluation,
repository snapshots, candidate-generation rules, framework adapters,
dependency-recovery logic, the \securityadg schema, representation
projections, and evaluation scripts remain fixed.
Graph construction uses neither human security labels nor model-generated
judgments, reproduction outcomes, vulnerability dispositions, or disclosure
status.

\paragraph{Use of generative AI tools.}
We used OpenAI Codex with the GPT-5.6 model as a supervised research
assistant during implementation and experimental execution. Its roles
included generating and revising analysis code, test scaffolding, experiment
orchestration commands, diagnostic suggestions, and figure-generation code.
The authors reviewed AI-assisted changes before their use in reported
experiments. We validated the resulting implementation through deterministic
unit tests, frozen-input reruns, schema and invariant checks, and recomputation
of reported aggregates from archived intermediate outputs.

We additionally used a frozen five-model panel to independently pre-annotate
150 candidate tasks and support disagreement-based triage. These outputs were
not used as ground-truth labels, were not consumed by \securityadg
construction, and did not determine reproduction or vulnerability
dispositions. Reference-set inclusion instead required inspectable evidence
from frozen source snapshots and controlled local reproduction reviewed by the
authors. The exact model identifiers, endpoints, and inference settings are
reported in the supplementary material.

\FloatBarrier
\subsection{Reproduction-Confirmed Ground Truth}
\label{sec:evaluation-ground-truth}
\label{sec:evaluation-disposition}

Static candidates do not constitute behavior or vulnerability ground truth.
We therefore establish a conservative reference set through targeted
reproduction. Validation progresses from source inspection to controlled
behavioral confirmation using, where appropriate, isolated execution,
stubbing or monkeypatching of external effects, or paired attack--defense
tests.

A case enters the reference set only if the relevant behavior can be tied
to the frozen source snapshot. Model assistance may support diagnostics or
test scaffolding, but model judgments are never treated as ground-truth
labels.

The resulting set contains 22 reproduction-confirmed behaviors across
13 repositories. Each case retains its frozen source location,
\securityadg instance, and recovered dependency path.

We conservatively assign each case one of three dispositions:

\begin{itemize}
    \item \textbf{Vulnerability ground truth}: reproduction and security
    evidence support a vulnerability conclusion under the evaluated threat
    model;

    \item \textbf{Pending disclosure candidate}: the behavior warrants
    further coordination or validation but is not treated as confirmed
    vulnerability ground truth; and

    \item \textbf{Guarded non-vulnerability}: the behavior is reproducible,
    but observed protections or usage constraints do not support a
    vulnerability conclusion under the tested conditions.
\end{itemize}

Table~\ref{tab:evaluation-overview} summarizes the reference set alongside
the corpus-wide analysis population.

\begin{table}[t]
\centering
\caption{Corpus-wide analysis and reproduction evidence. Candidate counts
are not vulnerability findings.}
\label{tab:evaluation-overview}
\small
\setlength{\tabcolsep}{5pt}
\renewcommand{\arraystretch}{1.10}
\begin{tabularx}{\linewidth}{@{}XrXr@{}}
\toprule
\multicolumn{2}{c}{\textbf{Corpus-Wide Analysis}} &
\multicolumn{2}{c}{\textbf{Reproduction Evidence}} \\
\cmidrule(r){1-2}
\cmidrule(l){3-4}
\textbf{Statistic} & \textbf{Value} &
\textbf{Outcome} & \textbf{Count} \\
\midrule
Frozen repositories & 67 & Reproduction-confirmed behaviors & 22 \\
Agent ecosystems & 11 & Repositories represented & 13 \\
Source files & 37,542 & Vulnerability ground truth & 1 \\
Source size & 291.2\,MB & Pending disclosure candidate & 1 \\
Candidate-bearing repositories & 65 & Guarded non-vulnerability & 20 \\
Candidate-bearing source files & 7,214 & & \\
Current full-corpus candidates & 23,866 & & \\
Generated \securityadg artifacts & 23,866 & & \\
\bottomrule
\end{tabularx}
\end{table}

The supplementary material documents the auxiliary 344-operation inventory
and matching protocol. Incomplete independent adjudication precludes aggregate
recovery metrics and exhaustive repository-wide ground-truth claims.

\subsection{RQ1: Corpus-Wide Coverage and Scalability}
\label{sec:results-rq1}

\paragraph{Question.}
\emph{How does \system behave when applied to the full \benchmark corpus in
terms of candidate scale, recovered security-context evidence, graph
construction success, and analysis cost?}

\paragraph{Results.}
The current scanner completes all 67 frozen repositories without skips or
failures and identifies 23,866 static security-sensitive operation
candidates across 65 repositories and 7,214 candidate-bearing source files.
\system emits one candidate-centered \securityadg artifact for every
candidate, with no reported construction failures. Following the operational
criterion in Section~\ref{sec:implementation-artifacts}, this measures
one-to-one generation and provenance linkage rather than semantic correctness.

Across this population, explicit framework-specific evidence is recovered
for 749 candidates (3.14\%), source-to-operation dependency evidence for
9,821 (41.15\%), and potential guard evidence for 3,075 (12.88\%).
Table~\ref{tab:corpus-evidence} summarizes these results. These frequencies
characterize evidence exposed by the current static analysis and do not
estimate vulnerability prevalence or repository-wide recall.

\begin{table}[t]
\centering
\caption{Evidence recovered in the current full-corpus run. Rates use all
23,866 candidates as the denominator.}
\label{tab:corpus-evidence}
\small
\setlength{\tabcolsep}{8pt}
\renewcommand{\arraystretch}{1.10}
\begin{tabular}{@{}lrr@{}}
\toprule
\textbf{Evidence or artifact} & \textbf{Count} & \textbf{Rate} \\
\midrule
Framework-specific evidence & 749 & 3.14\% \\
Source-to-operation dependency evidence & 9,821 & 41.15\% \\
Potential guard evidence & 3,075 & 12.88\% \\
Generated \securityadg artifacts & 23,866 & 100.0\% \\
\bottomrule
\end{tabular}
\end{table}

The exported corpus-wide aggregates give identical counts for source,
dependency, and trust-crossing fields. Because these aggregates do not
demonstrate three independent recovery events, we conservatively report one
source-to-operation dependency-evidence measure rather than triple-counting
the same 9,821 candidates. Independent trust-boundary recovery accuracy is
not measured by this experiment. The current extraction rules recover no
explicit privilege-context evidence in this run. Every candidate is assigned
an abstract external-effect node by its operation category; this is a graph
schema property rather than a recovered or runtime-confirmed effect.

Evidence recovery varies across operation categories. Source-to-operation
dependency evidence is recovered for 84.14\% of filesystem-write candidates
and 65.28\% of filesystem-read candidates, compared with 11.28\% of
external-tool invocations and 1.87\% of browser-control candidates.
Potential guard evidence ranges from 51.29\% for filesystem deletion and
41.82\% for dynamic-code execution to 1.37\% for external-tool invocation.
These differences characterize the current analyzer and corpus; they must
not be interpreted as differences in vulnerability prevalence. The complete
per-category table is provided in the supplementary material.

\paragraph{Scalability.}
Candidate extraction requires 1,508.5 seconds and \securityadg construction
1,538.2 seconds, for an end-to-end wall time of 3,046.6 seconds
(50.8 minutes) and throughput of 7.83 candidates/s. Combined per-repository
analysis time has a median of 9.6 seconds, with P90 and P95 values of 203.0
and 236.2 seconds, respectively. Table~\ref{tab:scalability} reports the
observed run. The latency distribution is strongly skewed toward larger
repositories.

\begin{table}[t]
\centering
\caption{Scalability results from one completed full-corpus run.}
\label{tab:scalability}
\small
\setlength{\tabcolsep}{10pt}
\renewcommand{\arraystretch}{1.10}
\begin{tabular}{@{}lr@{}}
\toprule
\textbf{Metric} & \textbf{Observed result} \\
\midrule
Repositories completed & 67/67 \\
Candidates / generated graphs & 23,866 / 23,866 \\
Candidate extraction & 1,508.5\,s \\
\securityadg construction & 1,538.2\,s \\
End-to-end wall time & 3,046.6\,s (50.8\,min) \\
Throughput & 7.83 candidates/s \\
Median repository time & 9.6\,s \\
P90 / P95 repository time & 203.0\,s / 236.2\,s \\
\bottomrule
\end{tabular}
\end{table}

\paragraph{Finding.}
\system processes the complete 67-repository corpus and emits
candidate-centered graph artifacts for all 23,866 detected operations without
reported construction failures. Security-context evidence is unevenly distributed:
41.15\% of candidates contain recovered source-to-operation dependency
evidence and 12.88\% expose potential guard evidence. The observed runtime
supports corpus-scale offline analysis, although large repositories produce
a pronounced latency tail.

\subsection{RQ2: Reproduction-Confirmed Evidence}
\label{sec:results-rq2}

\paragraph{Question.}
\emph{Can reproduction-backed validation establish conservative ground truth
for security-sensitive agent behaviors without relying on model judgments or
single-reviewer labels?}

\paragraph{Results.}
We reproduce 22 security-sensitive behaviors across 13 repositories.
Only one case currently satisfies our criteria for vulnerability ground
truth; one additional case remains pending disclosure, while 20 are guarded
behaviors that do not meet the vulnerability criteria under the evaluated
conditions.

This distribution shows that reproducing a sensitive behavior does not by
itself establish a vulnerability. Even when an agent-facing value
reaches a powerful capability, deployment exposure, authorization, privilege,
guards, and impact remain relevant to the vulnerability determination.

\paragraph{Finding.}
Reproduction-backed validation provides auditable behavior-level evidence
while preserving the distinction between security-sensitive behavior and
vulnerability. Neither static candidates nor model-generated judgments are
used as ground truth.

\subsection{RQ3: Beyond Sink Identity}
\label{sec:results-rq3}

\paragraph{Question.}
\emph{What security-relevant distinctions does \securityadg expose beyond
sensitive-operation identity alone?}

\paragraph{Results.}
Across the reproduction-confirmed cases, superficially similar sensitive
operations exhibit materially different security contexts. Recurring
patterns include direct versus removed source dependencies, guarded versus
unguarded execution, validation paths versus effectful execution paths, and identical
low-level operations exposed under different agent or trust contexts.

For example, an agent-facing parameter may directly influence command
execution in one case but be overwritten by a constant in another.
Similarly, an otherwise equivalent dependency path may be accompanied by an
allowlist, authentication check, fixed argument construction, path
restriction, or sandboxing mechanism.

A sink-only representation cannot express these distinctions because it
records only operation identity. \securityadg instead retains source
provenance, dependency, trust-boundary, guard, and effect evidence associated
with the same candidate.

\paragraph{Finding.}
Sensitive-operation identity is useful for candidate discovery but
insufficient for security interpretation. The additional dependency and
security-context evidence preserved by \securityadg distinguishes behaviors
that appear equivalent under an operation-only view.

\subsection{RQ4: Context Preservation}
\label{sec:results-rq4}
\label{sec:evaluation-baselines}
\label{sec:evaluation-metrics}

\paragraph{Question.}
\emph{How much security-relevant context does \securityadg preserve compared
with sink-only and simplified ADG representations?}

\paragraph{Setup.}
We derive three deterministic projections from the same underlying
candidate: a \emph{sink-only} view retaining only the sensitive operation;
a \emph{simplified ADG} retaining structural operation and external-effect
information; and the full \securityadg retaining source/dependency,
trust-boundary, guard, and external-effect evidence.

All three representations derive from the same candidate, so their
differences reflect information lost by the representation rather than
differences in candidate generation.

We evaluate nine evidence-backed held-out cases: five validated by controlled
local monkeypatch proofs and four by source-level construction proofs that did
not execute the external effect. Five cases contain reference guard evidence.
Before scoring, we freeze a case-level reference matrix from the corresponding
evidence records. Let $A_i$ contain the independently supported, applicable
context elements for case $i$, and let $R_i(v)\subseteq A_i$ be those retained
by representation $v$. Absent or unknown reference elements are excluded.
Context completeness is

\begin{equation}
\operatorname{CC}(v)
=
\frac{\sum_i |R_i(v)|}{\sum_i |A_i|},
\label{eq:context-completeness}
\end{equation}

An explicitly recorded \texttt{no\_guard\_confirmed} posture is treated as
a supported negative reference element rather than an absent or unknown one;
it therefore remains in the denominator, although a representation receives
positive guard-visibility credit only for concrete guard evidence.

The sums give a micro-average. \emph{Guard visibility} counts cases in
which both the reference and representation expose guard evidence. The
supplementary matrix reports each case's applicable elements, evidence
provenance, projection-specific retention, and score, making the 91.1\%
aggregate auditable.

\paragraph{Results.}
Table~\ref{tab:context-preservation} reports the comparison.

\begin{table}[H]
\centering
\caption{Security-context preservation across representation variants.
Higher values indicate that more reference context is preserved.}
\label{tab:context-preservation}
\small
\setlength{\tabcolsep}{10pt}
\renewcommand{\arraystretch}{1.12}
\begin{tabular}{@{}lcc@{}}
\toprule
\textbf{Representation}
& \textbf{Context Completeness}
& \textbf{Guard Visibility} \\
\midrule
Sink-only
& 20.0\%
& 0/5 \\

Simplified ADG
& 40.0\%
& 0/5 \\

\midrule
\textbf{\securityadg}
& \textbf{91.1\%}
& \textbf{5/5} \\
\bottomrule
\end{tabular}
\end{table}

The sink-only view preserves 20.0\% of the applicable reference context.
Adding structural operation and effect information increases completeness to
40.0\%, but does not expose any of the five observed guards.
The full \securityadg reaches 91.1\% context completeness and preserves all
five instances of guard evidence.

The increase from 40.0\% to 91.1\% shows that structural dependency
information alone does not account for the improvement; source provenance,
security-relevant dependency, trust-boundary, and guard evidence contribute
substantial additional context.

\paragraph{Finding.}
\securityadg preserves substantially more independently supported security
context than either reduced representation. These figures measure
\emph{context preservation}, not vulnerability-detection accuracy, and
guard visibility in all five cases does not imply that the observed guards
are sufficient.
\section{Discussion and Threats to Validity}
\label{sec:discussion}
\label{sec:threats}

\subsection{Interpreting ``Beyond Predefined Sinks''}
\label{sec:discussion-sinks}

Our results do not imply that predefined security-sensitive operations
should be discarded. Such specifications remain a practical and scalable
mechanism for locating code that warrants further analysis. The limitation is
that operation identity alone does not determine whether an operation is
agent-influenced, whether it crosses a trust boundary, or what protection
mechanisms constrain its use.

Accordingly, ``beyond predefined sinks'' refers to augmenting sink identity
rather than replacing sink specifications. \system uses predefined operations
as candidate anchors and then recovers agent semantics, source provenance,
dependency, trust-boundary, guard, and external-effect evidence for each one.

This distinction is reflected in the representation study: retaining only
operation identity preserves substantially less independently supported
security context than the full \securityadg. The benefit of \securityadg therefore
lies not in eliminating sink specifications, but in providing the
evidence needed to interpret superficially similar operations differently.

\subsection{From Security Evidence to Vulnerability}
\label{sec:discussion-vulnerability}

A reproduced security-sensitive behavior is not necessarily a vulnerability.
The reproduction study demonstrates this distinction directly: behaviors can
be tied to frozen source snapshots and exercised under controlled conditions
while still being constrained by validation, authorization, sandboxing,
fixed argument construction, deployment assumptions, or intended usage.

Static evidence should therefore be interpreted according to its strength.
A recovered dependency indicates that the available source-level evidence
supports a possible influence from an agent-relevant value to a sensitive
operation. A recovered guard indicates that potentially relevant defensive
evidence is visible. Neither observation alone proves exploitability,
vulnerability, or safety.

We therefore view \securityadg as an evidence representation rather than a
binary vulnerability classifier. Candidate-centered graphs support a staged
workflow: inexpensive static analysis identifies and contextualizes behaviors,
and costlier reproduction and security review examine cases that warrant it.

This separation also supports security auditing, candidate prioritization,
regression analysis, and responsible-disclosure triage, provided that these
uses preserve the distinction between evidence and vulnerability conclusions.

\subsection{Limitations and Threats to Validity}
\label{sec:discussion-limitations}

\paragraph{Construct and internal validity.}
Our evaluation relies on distinctions among sensitive operations,
security-critical behaviors, and vulnerabilities, as well as on the context
dimensions in \securityadg. These abstractions omit some deployment-dependent
factors. Context completeness measures represented evidence, not proven
reachability or exploitability; guard visibility does not establish sufficiency.

Errors in candidate rules, adapters, dependency recovery, or reproduction
could affect the results. We mitigate this risk by freezing snapshots and
analysis configurations before evaluation, separating graph construction
from labels, and retaining source locations, graphs, and dependency evidence.

The analysis is intentionally incomplete. Python recovery is local and
AST-based, not a sound whole-program, alias, points-to, or symbolic analysis.
TypeScript and JavaScript rely on weaker, syntax-aware lexical tracing.
Reflection, dynamic registration, metaprogramming, or deployment-specific
exposure and privilege may therefore yield missing or imprecise evidence.
Privilege context is represented only when explicit evidence is recoverable
from supported source-level constructs; the current full-corpus extraction
recovers none. In addition, the available corpus-wide aggregates report
identical source, dependency, and trust-crossing counts. We therefore do not
interpret them as three independent recovery measurements or claim an
independently validated trust-boundary recovery rate.

\paragraph{External validity.}
\benchmark contains 67 frozen repositories from 11 agent ecosystems but is
not statistically representative of the entire ecosystem. Availability,
selection criteria, and open-source access shape its composition. Python
dominates the sample, and \system analyzes Python dependencies more strongly
than TypeScript/JavaScript dependencies; aggregate results are therefore not
language-independent performance estimates.

\paragraph{Conclusion validity.}
The current full-corpus candidate population is generated by a later scanner
version than the frozen five-repository pilot manifest. The 23,866-candidate
full-corpus population and historical 531-candidate cohort are therefore not
directly comparable as a corpus-size scaling experiment. Re-running the
current scanner on the same five snapshots yields 508 candidates because
language scoping and AST filtering changed; we preserve the historical
manifest rather than merging the populations.

Corpus-wide evidence frequencies are descriptive properties of the current
analyzer and benchmark. They neither estimate vulnerability prevalence nor
establish recall with respect to all security-relevant behaviors in the
repositories. The reported timing comes from one completed run rather than a
repeated-run distribution, and the archived aggregate does not include peak
memory usage; scalability conclusions are correspondingly limited to the
observed run.

The five-model panel was used only to prioritize candidate inspection and did
not determine reference-set eligibility. Because this triage may nevertheless
influence which cases receive earlier manual attention, we do not treat the
reproduction set as an unbiased estimate of corpus-wide behavior prevalence.

The reference set contains 22 behaviors across 13 repositories; the
representation study contains nine held-out cases, including five with guard
evidence. These samples support the reported case- and representation-level
findings, but not exhaustive recall or general vulnerability-detection claims.

We therefore do not use precision, recall, or F1 as primary metrics for the
reference set. Incomplete adjudication likewise precludes matched counts or
aggregate recovery metrics for the 344-operation inventory.
Likewise, 91.1\% context completeness and 5/5 guard visibility describe the
evaluated cases rather than universal performance.
\section{Related Work}
\label{sec:related-work}

\subsection{Security Evaluation of LLM Agents}

The growing autonomy of LLM agents has motivated extensive work on their
attack surfaces. Prior studies demonstrate indirect prompt
injection~\cite{greshake2023indirect,liu2024promptinjection,injecagent2024}, evaluate
attacks and defenses in realistic agent environments~\cite{agentdojo2024},
and study unsafe agent--tool interactions~\cite{toolemu2024}. Broader
benchmarks examine security failures across prompts, tools, and
memory~\cite{asb2025}, harmful multi-step agent misuse~\cite{agentharm2025}, and
real-world software-security tasks~\cite{secbench2025}.

Recent work broadens the threat model beyond prompt injection to poisoned
agent memory~\cite{agentpoison2024}, web-agent compromise~\cite{wasp2025},
resource abuse~\cite{agentdos2026}, and MCP-specific
risks~\cite{mcplandscape2026,mcpsafetybench2026,mcptox2026,mpma2026}. A recent
systematization organizes the design, attack, and defense landscape of
agentic systems~\cite{agentsecuritysok2026}. Large-scale ecosystem evidence
also shows that third-party agent skills can carry deliberately malicious
instructions and helper code~\cite{maliciousskills2026}.

These studies primarily focus on attack generation, adversarial evaluation,
or vulnerability benchmarking. \system addresses a complementary problem:
recovering auditable dependency and security-context evidence through static
analysis of sensitive behavior in real-world agent source code.

\subsection{Program Analysis for Agent Security}

Traditional program analysis provides several foundations for
security-oriented dependency reasoning. Program Dependence Graphs (PDGs)
capture control and data dependencies~\cite{pdg1987}, while Code Property
Graphs integrate syntax, control flow, and dependence information into a
unified representation for vulnerability discovery~\cite{cpg2014}.
Declarative systems such as QL further demonstrate that expressive program
queries can scale to large codebases~\cite{ql2016}. Dynamic information-flow
tracking in TaintDroid~\cite{taintdroid2010} and static source--sink analyses
such as FlowDroid~\cite{flowdroid2014} and YASA~\cite{yasa2026} provide
complementary taint-oriented reasoning.

Recent work adapts these techniques to LLM-integrated software.
AgentFuzz~\cite{agentfuzz2025} dynamically searches for inputs that drive
agents toward sensitive operations. TaintP2X~\cite{taintp2x2026} statically
traces LLM-generated values toward sensitive sinks, while Agent
Audit~\cite{agentaudit2026} combines agent-aware data flow with credential,
configuration, and privilege analysis.

\system differs in its analytical objective. Rather than treating a predefined
source--sink relation as sufficient to characterize security risk, it uses
sensitive operations as candidate anchors and augments them with agent
semantics, source provenance, dependency, trust-boundary, guard, and
external-effect evidence.

\subsection{Security Enforcement for LLM Agents}

A complementary line of work aims to prevent unsafe agent behavior through
isolation, information-flow control, and least-privilege enforcement.
IsolateGPT~\cite{isolategpt2025} isolates LLM-integrated applications from
system resources, CaMeL~\cite{camel2026} separates trusted control flow from
untrusted data, Progent~\cite{progent2025} introduces programmable
least-privilege policies over tool calls, and ACE~\cite{ace2026} separates
planning from execution while enforcing information-flow and capability
boundaries. AttriGuard instead performs runtime causal attribution of tool
invocations to identify actions driven by untrusted
observations~\cite{attriguard2026}.

These systems primarily enforce security policies at runtime. \system is
complementary: it does not enforce a policy, but statically records evidence
about trust boundaries, guards, statically available privilege context, and
external effects so that existing agent software can be audited.

\subsection{Agent Dependency Representations and Positioning}

\begin{table}[H]
\centering
\caption{Positioning against closely related approaches.
``Context'' denotes explicit security-relevant evidence beyond operation
identity.}
\label{tab:related-positioning}
\small
\setlength{\tabcolsep}{5pt}
\renewcommand{\arraystretch}{1.08}
\begin{tabular}{@{}lcccl@{}}
\toprule
\textbf{Approach}
& \textbf{Setting}
& \textbf{Agent Dep.}
& \textbf{Context}
& \textbf{Primary Focus} \\
\midrule
AgentFuzz~\cite{agentfuzz2025}
& Grey-box & No & Partial & Vulnerability triggering \\

TaintP2X~\cite{taintp2x2026}
& Static & Partial & Partial & Prompt-to-sink detection \\

Agent Audit~\cite{agentaudit2026}
& Static & Partial & Yes & Agent security auditing \\

AgentFlow~\cite{agentflow2026}
& Static & Yes & Partial & Dependency recovery \\

Agent-BOM~\cite{agentbom2026}
& Runtime & Yes & Yes & Runtime reconstruction \\

\midrule
\textbf{\system}
& \textbf{Static}
& \textbf{Yes}
& \textbf{Yes}
& \textbf{Contextual behavior evidence} \\
\bottomrule
\end{tabular}
\end{table}

Agent frameworks encode relationships that conventional host-language
representations do not fully capture. AgentFlow~\cite{agentflow2026} builds
Agent Dependency Graphs over agents, prompts, models, capabilities, memory,
and control policies. Agent-BOM~\cite{agentbom2026} combines static
capabilities with runtime state to reconstruct security-relevant execution
and attack chains. \system also uses a graph representation, but at a
different granularity: the candidate-centered \securityadg augments one
sensitive operation with the dependencies and context needed to interpret it.

These systems are complementary: taint analyses recover source-to-operation
paths, agent-aware analyzers add security semantics, AgentFlow recovers
framework dependencies, and Agent-BOM reconstructs runtime context. \system
combines candidate-centered static recovery with explicit security context.
\section{Conclusion}
\label{sec:conclusion}

LLM-agent software connects model-generated decisions to security-sensitive
capabilities, making operation identity alone insufficient for interpreting
security-relevant behavior.

We presented \system, a static analysis framework that treats sensitive
operations as candidate anchors and constructs candidate-centered
\securityadg representations. By preserving agent relevance, provenance,
dependencies, trust boundaries, candidate guards, available privilege
information, and external effects, \system separates operation detection
from security interpretation.

We also introduced \benchmark, a reproducible corpus of 67 frozen
repositories spanning 11 ecosystems and 37,542 source files. Applying the
current analyzer to the full corpus yields 23,866 static security-sensitive
operation candidates across 65 repositories, for which \system emits
23,866 candidate-centered \securityadg artifacts without reported
construction failures. Corpus-wide analysis
recovers source-to-operation dependency evidence for 41.15\% of candidates
and potential guard evidence for 12.88\%, completing in 50.8 minutes.

A separate reproduction-backed evaluation establishes a conservative
reference set of 22 behaviors: one confirmed vulnerability, one pending
disclosure candidate, and 20 guarded behaviors that did not meet the
vulnerability criteria.

In the held-out study, \securityadg preserves 91.1\% of the reference context
and all five observed guards, compared with 20.0\% for a sink-only view and
40.0\% for a simplified ADG; neither reduced view exposes the guards.
These results support the central conclusion of this work:
predefined sensitive operations remain useful analysis anchors, but
security-critical behavior in LLM-agent software is interpreted more
faithfully when operation identity is augmented with dependency and
security-context evidence.

\section*{Data and Artifact Availability}

The implementation of \system, the analysis pipeline, experiment
configuration, evaluation scripts, and documentation are publicly available
at \url{https://github.com/A1LinLin1/AgentSecBench}. The repository is released
under the Apache License 2.0. Security-sensitive reproduction details
associated with ongoing coordinated disclosures are intentionally withheld.
They will be released, where appropriate, after remediation or coordinated
public disclosure.

% TODO(arXiv): Add an immutable GitHub Release or Zenodo DOI before submission.
% TODO(arXiv): Add acknowledgments, funding, and conflict disclosures if applicable.

\bibliographystyle{ACM-Reference-Format}
\bibliography{references}

\clearpage
\appendix
\section{AgentSecBench Repository Inventory and Collection Criteria}
\label{sec:supp-corpus-inventory}

\benchmark contains 67 publicly available LLM-agent repositories spanning 11
agent ecosystems. Corpus construction was performed independently of
\system output. We retained repositories with substantive agent logic and
analyzable source code, including MCP-based systems, coding agents,
framework-based applications, autonomous agents, and multi-agent platforms.
We excluded duplicates, mirrors, trivial examples, and repositories unsuitable
for reproducible source-level analysis.

Each retained repository is identified by its repository coordinate and an
immutable Git commit. Candidate extraction, graph construction, reproduction,
and evaluation use that same frozen snapshot. The machine-readable frozen
corpus manifest in the \ifarxiv public artifact repository\else anonymous
replication package\fi{} provides the complete
row-level repository inventory and commit identifiers; the PDF reports the
aggregate composition in Table~\ref{tab:supp-corpus-composition} rather than
duplicating the long manifest.

\begin{table}[t]
\centering
\caption{Aggregate composition of the frozen \benchmark corpus. Candidate
counts characterize the current analyzer and are not vulnerability counts.}
\label{tab:supp-corpus-composition}
\small
\setlength{\tabcolsep}{7pt}
\renewcommand{\arraystretch}{1.08}
\begin{tabular}{@{}lr@{}}
\toprule
\textbf{Corpus property} & \textbf{Value} \\
\midrule
Frozen repositories & 67 \\
Agent ecosystems & 11 \\
Source files & 37,542 \\
Source size & 291.2\,MB \\
Primarily Python repositories & 51 \\
Primarily TypeScript repositories & 7 \\
Repositories in other primary languages & 9 \\
Repositories with current-scan candidates & 65 \\
Current full-corpus candidates & 23,866 \\
\bottomrule
\end{tabular}
\end{table}

The nine repositories in the remaining-language group primarily use C\#,
Rust, Go, PHP, or Shell. Repository inclusion is not conditioned on candidate
counts, graph construction outcomes, reproduction outcomes, vulnerability
dispositions, or model-generated judgments. The frozen manifest is therefore
an input to the evaluation rather than an output selected by the analyzer.
\section{Framework Normalization and Matching Rules}
\label{sec:supp-framework-normalization}

Agent frameworks expose tools and capabilities through heterogeneous source
constructs. The adapter layer maps a matched construct to the normalized
record

\begin{equation}
\mathcal{N}=\langle F,R_s,Y,P_s,Q\rangle,
\label{eq:supp-framework-normalization}
\end{equation}

where $F$ is the framework family, $R_s$ is the semantic role, $Y$ is the
associated callable or program symbol, $P_s$ is the set of exposed parameters,
and $Q$ records the matching evidence and confidence. Normalization determines
how source constructs are represented; it does not assign vulnerability
labels.

\begin{table}[t]
\centering
\caption{Framework construct families represented by the frozen adapter
registry. Concrete selectors are versioned with the implementation.}
\label{tab:supp-framework-patterns}
\small
\setlength{\tabcolsep}{4pt}
\renewcommand{\arraystretch}{1.08}
\begin{tabularx}{\linewidth}{@{}l>{\raggedright\arraybackslash}X>{\raggedright\arraybackslash}X@{}}
\toprule
\textbf{Framework family} & \textbf{Construct family} &
\textbf{Normalized interpretation} \\
\midrule
MCP / FastMCP & Tool decorators and server-side tool registration &
Agent-facing callable and exposed parameters \\
LangChain & Tool decorators, callable wrappers, and toolkit exports &
Tool capability associated with its callable \\
CrewAI & Tool declarations and tool collections attached to agents or tasks &
Agent-available capability \\
AutoGen & Function or tool registration between conversational agents and executors &
Registered callable capability \\
OpenAI Agents SDK & Function-tool wrappers and tools attached to an agent &
Agent-facing callable capability \\
Semantic Kernel & Kernel-function declarations and plugin methods &
Kernel-exposed callable capability \\
LlamaIndex & Function/query tool wrappers and tool collections &
Agent- or query-facing callable capability \\
\bottomrule
\end{tabularx}
\end{table}

Adapters apply frozen framework-specific selectors to decorators, calls,
registrations, wrappers, workflow nodes, or tool classes. When available,
module qualification and imported aliases are normalized before matching. A
successful match records the construct form, associated symbol, source
location, exposed parameters, and framework provenance. Ambiguous or
unsupported constructs remain available to the language-level analysis but
are not promoted to framework-specific evidence solely from a security label,
model judgment, or reproduction outcome.

Framework normalization is independent of candidate-generation and graph
schema decisions. Adding an adapter extends semantic recognition without
changing the candidate-centered \securityadg representation. The frozen
adapter registry and its tests are part of the complete method implementation
\ifarxiv available in the public artifact repository\else to be included in
the anonymous replication package\fi{}.
\section{Generative-AI-Assisted Candidate Triage}

We evaluated a frozen panel of five model raters over 150 candidate tasks.
Each rater processed every task independently in a fresh single-turn request.
Raters did not receive scanner confidence, detector identity, sampling weight,
human annotations, or another model's response. Tools, browsing, retrieval,
memory, and code execution were disabled. Responses were required to follow a
fixed JSON schema and were validated locally. Transport and schema failures
were recorded rather than silently discarded.

Four raters were accessed through OpenRouter using the exact routing slugs
\nolinkurl{openai/gpt-5-mini},
\nolinkurl{anthropic/claude-sonnet-4.6},
\nolinkurl{google/gemini-3.5-flash-lite}, and
\nolinkurl{qwen/qwen3-max}. Provider fallbacks were disabled, parameter support
was required, and the run recorded the returned model identifier, generation
identifier, upstream provider, token usage, cost, and system fingerprint. The
fifth rater, \nolinkurl{deepseek-v4-pro}, was accessed through the official
DeepSeek API rather than OpenRouter. The completed panel contains 750 valid
task--rater responses (150 tasks times five raters).

The panel was used only for pre-annotation and disagreement-based triage.
Inter-model agreement was not interpreted as human inter-annotator agreement
or as evidence of semantic correctness. Panel responses were excluded from
\securityadg construction and did not establish behavior or vulnerability
ground truth.

\begin{table}[t]
\centering
\caption{Frozen five-model pre-annotation panel.}
\label{tab:model-panel}
\small
\setlength{\tabcolsep}{4pt}
\renewcommand{\arraystretch}{1.10}
\begin{tabularx}{\linewidth}{@{}l>{\raggedright\arraybackslash}X>{\raggedright\arraybackslash}X@{}}
\toprule
\textbf{Rater} & \textbf{Exact identifier and interface} &
\textbf{Inference and output settings} \\
\midrule
M1 & \nolinkurl{openai/gpt-5-mini}; OpenRouter &
Medium reasoning; reasoning excluded; JSON Schema \\
M2 & \nolinkurl{anthropic/claude-sonnet-4.6}; OpenRouter &
Medium reasoning; reasoning excluded; JSON Schema \\
M3 & \nolinkurl{google/gemini-3.5-flash-lite}; OpenRouter &
Medium reasoning; reasoning excluded; JSON Schema \\
M4 & \nolinkurl{qwen/qwen3-max}; OpenRouter &
Provider-default reasoning; JSON Schema \\
M5 & \nolinkurl{deepseek-v4-pro}; DeepSeek official API &
Thinking enabled, high; reasoning not stored; JSON object with local schema
validation \\
\bottomrule
\end{tabularx}
\end{table}

For M1--M4, requests used
\url{https://openrouter.ai/api/v1/chat/completions} with
\nolinkurl{allow_fallbacks=false}, \nolinkurl{require_parameters=true}, and
\nolinkurl{data_collection=deny}. Zero-data-retention endpoints were preferred,
but the recorded run explicitly allowed non-ZDR endpoints under the
data-collection-denial policy. The recorded upstream providers were OpenAI,
Anthropic, Google AI Studio, and Alibaba, respectively, and provider
consistency checks passed. For M5, requests used
\url{https://api.deepseek.com/chat/completions} with a separate API
credential. Because DeepSeek's JSON mode did not enforce the supplied JSON
Schema, its outputs underwent the same local schema validation as the other
responses. No API keys, account identifiers, billing records, or
author-identifying generation URLs are included in the artifact.

All model-generated outputs used in the study were validated by the authors
or by deterministic validation scripts. Model judgments were not treated as
vulnerability ground truth. The authors remain fully responsible for the
study design, implementation, analysis, and claims.
\section{Held-Out Context-Preservation Study}

We evaluate three representation projections over nine held-out
evidence-backed cases: five validated by controlled local monkeypatch proofs
and four by source-level construction proofs that did not execute the external
effect. Each case-level reference contains independently supported context
dimensions for operation, external effect, dependency path, trust boundary,
and guard posture. The four cases recorded as
\texttt{no\_guard\_confirmed} contain an explicit negative guard-posture
observation rather than an absent or unknown field; the guard-posture dimension
therefore remains in the denominator. Positive guard-visibility credit is
awarded only when concrete guard evidence is exposed by the representation.
These measurements concern representation retention, not
vulnerability-detection accuracy.

\begin{table*}[t]
\caption{Evidence provenance for the held-out context-preservation cases.
Repository identifiers are retained to make the frozen evidence auditable; no
row is a vulnerability claim.}
\label{tab:supp-context-provenance}
\centering
\small
\setlength{\tabcolsep}{3pt}
\begin{tabularx}{\textwidth}{@{}l>{\raggedright\arraybackslash}p{3.2cm}>{\raggedright\arraybackslash}X>{\raggedright\arraybackslash}p{2.7cm}>{\raggedright\arraybackslash}p{2.7cm}@{}}
\toprule
Case & Repository & Behavior & Evidence mode & Guard posture \\
\midrule
HSQ-0014 & \nolinkurl{dddabtc/winremote-mcp} & shell command execution & local monkeypatch proof & \nolinkurl{no_guard_confirmed} \\
HSQ-0087 & \nolinkurl{dddabtc/winremote-mcp} & application launch & local monkeypatch proof & \nolinkurl{no_guard_confirmed} \\
HSQ-0089 & \nolinkurl{dddabtc/winremote-mcp} & desktop notification script & local monkeypatch proof & guarded \\
HSQ-0091 & \nolinkurl{dddabtc/winremote-mcp} & network diagnostic process & local monkeypatch proof & \nolinkurl{no_guard_confirmed} \\
HSQ-0103 & \nolinkurl{dddabtc/winremote-mcp} & service control & local monkeypatch proof & guarded \\
HSQ-0211 & \nolinkurl{nexu-io/html-anything} & agent CLI discovery process & source-level construction proof & \nolinkurl{no_guard_confirmed} \\
HSQ-0298 & \nolinkurl{nexu-io/html-anything} & external content file write & source-level construction proof & guarded \\
HSQ-0300 & \nolinkurl{nexu-io/html-anything} & external tarball extraction & source-level construction proof & guarded \\
HSQ-0302 & \nolinkurl{nexu-io/html-anything} & validated skill file write & source-level construction proof & guarded \\
\bottomrule
\end{tabularx}
\end{table*}

\begin{table*}[t]
\caption{Case-level representation retention. Each case has five applicable
reference dimensions: operation, effect, dependency, trust boundary, and guard
posture.}
\label{tab:supp-context-cases}
\centering
\small
\begin{tabular}{llrrrr}
\toprule
Case & Guard posture & Applicable & Sink-only & Simplified ADG & Security-ADG \\
\midrule
HSQ-0014 & no\_guard\_confirmed & 5 & 1/5 & 2/5 & 4/5 \\
HSQ-0087 & no\_guard\_confirmed & 5 & 1/5 & 2/5 & 4/5 \\
HSQ-0089 & guarded & 5 & 1/5 & 2/5 & 5/5 \\
HSQ-0091 & no\_guard\_confirmed & 5 & 1/5 & 2/5 & 4/5 \\
HSQ-0103 & guarded & 5 & 1/5 & 2/5 & 5/5 \\
HSQ-0211 & no\_guard\_confirmed & 5 & 1/5 & 2/5 & 4/5 \\
HSQ-0298 & guarded & 5 & 1/5 & 2/5 & 5/5 \\
HSQ-0300 & guarded & 5 & 1/5 & 2/5 & 5/5 \\
HSQ-0302 & guarded & 5 & 1/5 & 2/5 & 5/5 \\
\midrule
Micro total & -- & 45 & 9/45 & 18/45 & 41/45 \\
\bottomrule
\end{tabular}
\end{table*}

\begin{table}[t]
\caption{Micro-averaged context completeness and positive guard visibility.}
\label{tab:supp-context-summary}
\centering
\small
\begin{tabular}{lrrr}
\toprule
Representation & Retained & Completeness & Guard visibility \\
\midrule
Sink-only & 9/45 & 20.0\% & 0/5 \\
Simplified ADG & 18/45 & 40.0\% & 0/5 \\
Security-ADG & 41/45 & 91.1\% & 5/5 \\
\bottomrule
\end{tabular}
\end{table}

The micro totals reproduce the values reported in the main paper:
$9/45=20.0\%$, $18/45=40.0\%$, and $41/45=91.1\%$. \securityadg exposes
concrete guard evidence in all five guarded cases. The four remaining
reference records explicitly report \texttt{no\_guard\_confirmed}; because the
graph does not emit a positive guard node for these cases, they contribute no
guard-retention credit.
\section{Corpus-Wide Characterization and Scalability}
\label{sec:supp-corpuswide}

This section provides the detailed corpus-wide characterization underlying
RQ1 in the main paper. All measurements in this section use the
\emph{current full-corpus run}: the current candidate extractor applied to
all 67 frozen repositories in \benchmark. This population is distinct from
the historical 531-candidate frozen pilot cohort and the 508 candidates
obtained by re-running the current scanner on the same five pilot
repositories.

The current full-corpus run identifies 23,866 static security-sensitive
operation candidates from 7,214 source files. Sixty-five of the 67
repositories contain at least one candidate. All 67 repositories complete
candidate extraction and graph construction without reported skips or
failures. Candidate counts in this section are analysis populations and must
not be interpreted as vulnerability counts.

\subsection{Aggregate Security-ADG Evidence}

Table~\ref{tab:supp-corpus-evidence} summarizes the evidence exposed by the
current full-corpus analysis. For every candidate, \system emits one
candidate-centered \securityadg artifact linked to its candidate identifier
and source provenance.

\begin{table}[t]
\centering
\caption{Aggregate evidence recovered in the current full-corpus run. Rates
use all 23,866 static candidates as the denominator. Candidate counts are not
vulnerability findings.}
\label{tab:supp-corpus-evidence}
\small
\setlength{\tabcolsep}{8pt}
\renewcommand{\arraystretch}{1.08}
\begin{tabular}{@{}lrr@{}}
\toprule
\textbf{Evidence or artifact} & \textbf{Count} & \textbf{Rate} \\
\midrule
Framework-specific evidence & 749 & 3.14\% \\
Source-to-operation dependency evidence & 9,821 & 41.15\% \\
Potential guard evidence & 3,075 & 12.88\% \\
Generated \securityadg artifacts & 23,866 & 100.00\% \\
\bottomrule
\end{tabular}
\end{table}

The exported corpus-wide aggregates contain identical candidate counts for
source provenance, dependency paths, and trust-crossing fields: 9,821
candidates in each case. These exported counts do not establish that the
three fields were recovered by independent analytical events. We therefore
report the consolidated \emph{source-to-operation dependency evidence}
measure in Table~\ref{tab:supp-corpus-evidence} rather than treating the three
identical counts as independent results.

The current extraction rules recover no explicit privilege-context evidence
in this run. This does not imply that the analyzed programs execute without
privilege; rather, \system records privilege context only when it can be
derived from supported source-level constructs. Likewise, every candidate
receives an abstract external-effect node based on its operation category.
The resulting 100\% external-effect-node coverage is therefore a schema
property rather than evidence that the modeled effect occurred at runtime.

\subsection{Evidence by Operation Category}

Table~\ref{tab:supp-category-evidence} reports candidate counts and the
frequency of recovered dependency, guard, and trust-crossing fields for each
operation category. The category-level percentages characterize the current
analyzer and corpus only. They do not estimate vulnerability prevalence or
category-specific security risk.

\begin{table}[t]
\centering
\caption{Corpus-wide evidence by security-sensitive operation category.
Percentages are relative to the number of candidates in each row.}
\label{tab:supp-category-evidence}
\scriptsize
\setlength{\tabcolsep}{3.5pt}
\renewcommand{\arraystretch}{1.08}
\begin{tabular}{@{}lrrrr@{}}
\toprule
\textbf{Category} & \textbf{Candidates} & \textbf{Dependency} &
\textbf{Guard} & \textbf{Trust-crossing} \\
\midrule
browser\_control & 588 & 11 (1.87\%) & 13 (2.21\%) & 11 (1.87\%) \\
command\_execution & 2,722 & 933 (34.28\%) & 315 (11.57\%) & 933 (34.28\%) \\
credential\_access & 2,548 & 251 (9.85\%) & 454 (17.82\%) & 251 (9.85\%) \\
database\_access & 1,304 & 106 (8.13\%) & 36 (2.76\%) & 106 (8.13\%) \\
dynamic\_code\_execution & 55 & 24 (43.64\%) & 23 (41.82\%) & 24 (43.64\%) \\
external\_tool\_invocation & 5,344 & 603 (11.28\%) & 73 (1.37\%) & 603 (11.28\%) \\
filesystem\_delete & 583 & 235 (40.31\%) & 299 (51.29\%) & 235 (40.31\%) \\
filesystem\_read & 4,182 & 2,730 (65.28\%) & 1,062 (25.39\%) & 2,730 (65.28\%) \\
filesystem\_write & 5,013 & 4,218 (84.14\%) & 632 (12.61\%) & 4,218 (84.14\%) \\
message\_or\_email\_send & 164 & 70 (42.68\%) & 32 (19.51\%) & 70 (42.68\%) \\
network\_access & 1,157 & 553 (47.80\%) & 121 (10.46\%) & 553 (47.80\%) \\
permission\_or\_auth\_change & 206 & 87 (42.23\%) & 15 (7.28\%) & 87 (42.23\%) \\
\midrule
\textbf{Total} & \textbf{23,866} & \textbf{9,821 (41.15\%)} &
\textbf{3,075 (12.88\%)} & \textbf{9,821 (41.15\%)} \\
\bottomrule
\end{tabular}
\end{table}

Dependency evidence is strongly category-dependent. It is recovered for
84.14\% of filesystem-write candidates and 65.28\% of filesystem-read
candidates, compared with 11.28\% of external-tool invocations and 1.87\% of
browser-control candidates. Guard evidence is likewise uneven: 51.29\% of
filesystem-delete candidates and 41.82\% of dynamic-code-execution candidates
contain potential guard evidence, whereas only 1.37\% of external-tool
invocations do.

These differences should not be interpreted as evidence that one operation
category is more vulnerable than another. They reflect the interaction among
the current source-analysis rules, the coding patterns in the frozen corpus,
and the evidence that can be recovered statically.

As noted above, dependency and trust-crossing counts are identical in the
exported corpus-wide aggregate. The trust-crossing column is retained in
Table~\ref{tab:supp-category-evidence} for auditability of the archived
results, but we do not interpret it as an independently validated
trust-boundary recovery rate.

\subsection{Scalability}

The full-corpus timing experiment was executed using one analysis process
with no worker pool. Candidate extraction and graph construction were timed
without \texttt{tracemalloc} so that profiler overhead did not inflate the
reported wall-clock measurements.

Table~\ref{tab:supp-scalability} summarizes the completed run.

\begin{table}[t]
\centering
\caption{Scalability measurements from one completed full-corpus run. The
measurements characterize the observed execution environment and are not
presented as cross-system performance estimates.}
\label{tab:supp-scalability}
\small
\setlength{\tabcolsep}{7pt}
\renewcommand{\arraystretch}{1.08}
\begin{tabular}{@{}lr@{}}
\toprule
\textbf{Metric} & \textbf{Observed result} \\
\midrule
Frozen repositories processed & 67 / 67 \\
Repositories with candidates & 65 \\
Candidate-bearing source files & 7,214 \\
Static candidates & 23,866 \\
Generated \securityadg artifacts & 23,866 \\
Candidate extraction time & 1,508.456 s \\
Candidate extraction throughput & 15.821 candidates/s \\
Graph construction time & 1,538.164 s \\
Graph construction throughput & 15.516 candidates/s \\
End-to-end wall time & 3,046.620 s (50.78 min) \\
End-to-end throughput & 7.834 candidates/s \\
Median repository time & 9.562 s \\
P90 repository time & 203.008 s \\
P95 repository time & 236.191 s \\
\bottomrule
\end{tabular}
\end{table}

The per-repository timing distribution is strongly skewed. Although the
median combined analysis time is 9.562 seconds, P90 and P95 increase to
203.008 and 236.191 seconds, respectively. Reporting these quantiles is
important because an arithmetic mean alone would obscure the large-repository
tail.

The completed timing run predates formal process-level peak-RSS collection.
We therefore do not report a peak-memory result for this run. A separate
earlier \texttt{tracemalloc} experiment on the frozen 531-candidate pilot
cohort measured Python allocation behavior, but its profiler-inflated wall
time is not comparable with the ordinary full-corpus timing experiment and is
not used in the main scalability result.

\subsection{Population and Reproducibility Notes}

The full-corpus measurements above must not be compared directly with the
historical 531-candidate frozen pilot cohort as though repository count were
the only changing variable. The pilot manifest was generated by an earlier
scanner configuration. Applying the current scanner to the same five frozen
repository snapshots yields 508 candidates because subsequent language
scoping and AST filtering changed candidate extraction.

We therefore preserve the three populations separately:

\begin{itemize}
\item \textbf{Frozen pilot cohort:} 531 candidates from five repositories,
retained for compatibility with historical pilot and annotation studies;
\item \textbf{Current five-repository scan:} 508 candidates on the same
repository snapshots, used only to document scanner-version drift; and
\item \textbf{Current full-corpus run:} 23,866 candidates from all 67 frozen
repositories, used for the corpus-wide characterization and scalability
results reported above.
\end{itemize}

None of these candidate populations constitutes a vulnerability count, and
this descriptive corpus-wide experiment does not establish repository-wide
recall.
\section{Inventory-Bounded Evaluation}
\label{sec:supp-inventory}

The main evaluation uses corpus-wide descriptive measurements,
reproduction-confirmed behavior evidence, and a held-out
representation-preservation study. We additionally retain an auxiliary
source-level operation inventory to provide bounded evidence about operation
recovery on code that was separated from development-time method tuning.

This evaluation is deliberately described as \emph{inventory-bounded}. The
finite inventory does not establish that every relevant operation in either
repository has been enumerated and therefore cannot support claims of
repository-wide exhaustive recall.

\subsection{Held-Out Operation Inventory}

The frozen inventory contains 344 source-level operation records from two
repositories:

\begin{itemize}
\item \nolinkurl{dddabtc/winremote-mcp}: 148 operation records; and
\item \nolinkurl{nexu-io/html-anything}: 196 operation records.
\end{itemize}

The two repositories were held separate from development-time method tuning
when this inventory was constructed. The inventory covers eight
security-sensitive operation categories.

\begin{table}[t]
\centering
\caption{Category distribution of the frozen 344-operation held-out
inventory. The inventory is finite and is not treated as exhaustive
repository-wide ground truth.}
\label{tab:supp-inventory-categories}
\small
\setlength{\tabcolsep}{8pt}
\renewcommand{\arraystretch}{1.08}
\begin{tabular}{@{}lr@{}}
\toprule
\textbf{Category} & \textbf{Count} \\
\midrule
Command execution & 76 \\
Filesystem write & 62 \\
External tool invocation & 58 \\
Network access & 46 \\
Filesystem read & 43 \\
Browser control & 27 \\
Dynamic code execution & 17 \\
Filesystem delete & 15 \\
\midrule
\textbf{Total} & \textbf{344} \\
\bottomrule
\end{tabular}
\end{table}

The category distribution is not intended to estimate the prevalence of
security-sensitive operations in the broader LLM-agent ecosystem. It
describes only the frozen auxiliary inventory used for this bounded
evaluation.

\subsection{Reference Evidence}

Inventory entries are tied to frozen source locations and are reviewed using
inspectable source-level evidence. Where stronger validation is available for
a corresponding behavior, the evaluation distinguishes among the following
evidence levels:

\begin{itemize}
\item \textbf{Source-level evidence}, establishing that a relevant program
construction or dependency exists in the frozen source snapshot;
\item \textbf{Stub or monkeypatch evidence}, establishing which call or
argument is produced under a controlled input;
\item \textbf{Isolated execution evidence}, establishing that the
corresponding external effect occurs in a controlled environment; and
\item \textbf{Attack--defense paired evidence}, evaluating whether a
candidate protection mechanism constrains the corresponding behavior.
\end{itemize}

These evidence levels are not interchangeable. In particular, the
344-operation inventory primarily serves as a source-level operation
inventory and is not itself a set of confirmed vulnerabilities.
Model-generated judgments are not treated as reference labels.

\subsection{Prediction-to-Reference Matching}

Method outputs are matched to inventory entries using the frozen matching
rule

\begin{equation}
\mathrm{match}
=
\mathrm{repository}
+
\mathrm{commit}
+
\mathrm{file}
+
\mathrm{operation\ category}
+
\mathrm{line\ overlap}.
\label{eq:supp-inventory-match}
\end{equation}

No label-dependent line tolerance or manual per-item matching is introduced
after evaluation. A reference inventory record is considered recovered only
when a method output satisfies the frozen matching rule.

Let $I$ denote the 344-operation reference inventory and let $M$ denote the
set of method outputs matched to records in $I$. We define inventory-bounded
recall as

\begin{equation}
R_{\mathrm{inv}}
=
\frac{|M|}{|I|}.
\label{eq:supp-inventory-recall}
\end{equation}

Because $I$ is a finite reviewed inventory rather than an exhaustive
enumeration of all possible security-relevant operations in the two
repositories, $R_{\mathrm{inv}}$ must not be interpreted as repository-wide
recall.

\subsection{Evaluation Status and Claim Boundary}

The frozen inventory defines the finite population and the coordinate-level
matching protocol used by the evaluator. At the time of this study, however,
the independently adjudicated label file does not cover all 344 inventory
items. The evaluator therefore refuses to emit aggregate or category-level
recovery metrics, and we do not report matched counts, recall, precision, or
F1 for this population. Results from the smaller reviewed sample are not
substituted for full-inventory measurements. The inventory is retained here
to document the evaluation population and protocol; completing the
independent adjudication remains future work.

\subsection{Interpretation and Limitations}

This auxiliary evaluation serves a narrower purpose than the corpus-wide
characterization in Section~\ref{sec:supp-corpuswide}. The latter measures how
the current analyzer behaves over all 67 frozen repositories, whereas the
344-operation inventory provides a finite source-level denominator on two
held-out repositories.

The inventory therefore supports statements about recovery relative to its
reviewed records, but not claims of exhaustive repository-wide recall.
Likewise, an operation matched in this evaluation is not necessarily
agent-controlled, exploitable, or vulnerable. Behavior-level and
vulnerability conclusions remain governed by the separate
reproduction-backed evidence layer described in the main paper.

\end{document}